\documentclass[12pt]{article}
\title{Quantum Geometrodynamics of the Bianchi IX Cosmological Model}

\author{Arkady Kheyfets\thanks{Department of Mathematics, North Carolina
State University, Raleigh, NC 27695}, Warner A.
Miller\thanks{Department of Physics, Florida Atlantic University,
Boca Raton, FL 33431} and Ruslan Vaulin${}^\dag$}

\begin{document}

\maketitle
\begin{abstract}
 The canonical quantum theory of gravity -- Quantum Geometrodynamics
 (QG) is applied to the homogeneous Bianchi type IX cosmological
 model. As a result, the framework for the quantum theory of
 homogeneous cosmologies is developed.  We show that the theory is
 internally consistent, and prove that it possesses the correct
 classical limit (the theory of general relativity). To emphasize the
 special role that the constraints play in this new theory we, compare it
 to the traditional ADM square-root and Wheeler-DeWitt quantization
 schemes.  We show that, unlike the traditional approaches, QG leads
 to a well-defined Schr\"odinger equation for the wave-function of the
 universe that is inherently coupled to the expectation value of the constraint
 equations. This coupling to the constraints is responsible for the
 appearance of a coherent spacetime picture. Thus, the physical meaning
 of the constraints of the theory is quite different from Dirac's
 interpretation. In light of this distinctive feature of the theory,
 we readdress the question of the dark energy effects in the Bianchi
 IX cosmological model for highly non-classical quantum states. We
 show that, at least for this model, for any choice of the initial
 wave function, the quantum corrections will not produce the
 accelerated expansion of the universe.

\end{abstract}
\section{Introduction}

The canonical method of quantization is one of the most direct paths
to the quantum theory. It is appealing because it is based on the
well-established classical theory, and therefore leads to a
plausible physical theory by the virtue of Erenfest's
theorem. Therefore, the resulting quantum theory automatically
reproduces, at some limit, the observable physical effects described by
the classical theory. This feature is true for the general quantum
gravitational field.

It has been a long-standing problem to apply this method to Einstein's
geometric theory of gravity -- the theory of general relativity.  The
challenge is to find an appropriate time-translation operator, a
Hamiltonian of the theory, which would govern the dynamics of the
theory at the quantum level. The problem with general relativity is
that the Hamiltonian is the generator of gauge transformations and
therefore must vanish. Simply speaking, it is a constraint of the
theory. This does not cause any problems for the classical dynamics,
since one can still calculate the Poisson brackets using this
Hamiltonian and obtain Einstein's equations of motion. However, at the
quantum level, according to Dirac's treatment of the constraints, the
operator corresponding to the classical expression for the constraint
must annihilate physical states of the theory. This means that
classical Hamiltonian when expressed as a quantum operator fails to
generate time translations. A challenge facing canonical quantum
gravity is to find an alternative operator for time translation, this
is often referred to as ``problem of time.''
\cite{Kuc92,Kuc93,Isham99}.

In this article we apply the formalism of Quantum Geometrodynamics
(QG) \cite{KheMil94a,KheMil96,KheMil00,KheMil04} to the anisotropic
homogeneous Bianchi type IX cosmological model. Because of its
structure, it can be parameterized with only few degrees of freedom
instead of the infinity of degrees of freedom of the field in the
general case. On the other hand, due to its anisotropy, the model is
still rich enough to reflect some of the characteristic features of
the cosmological singularity,\cite{BK69,BKL71} and thus may be an
interesting candidate to test the quantum theory.

We have two goals in applying the general formalism to this concrete
system. First, we would like to emphasize the special role played by
the constraints in this theory, which is conceptually and
interpretationally different from the more traditional approaches.
Second, we wish to develop a quantum cosmology theory, within which,
one can address the issues relevant for cosmology in general. In
Sec.~2 of the article, we briefly review the general theory ( for
extended discussion see \cite{KheMil04}) mostly for the sake of
introducing notations. In Sec.~3 we introduce the Bianchi IX
cosmological model as a homogeneous solution of Einstein's equations
of motion. In the next two sections, Sec.~4 and 5, we apply the ADM
square-root quantization and Wheeler-DeWitt theory to the model, with
the aim to illustrate the relative successes and failures of these
formalisms.  We place special emphasis on the treatment of the
constraints in these theories.  In Sec.~6, we develop the quantum
theory of the Bianchi IX cosmological model within the general
formalism of QG. Here we show that, following the prescription, it is
possible to define a Hamiltonian operator which leads to a
Schr\"odinger equation coupled to the constraint equations. We show,
quite generally, that the resulting quantum theory reproduces
Einstein's equations in its classical limit.  The detailed proof of
this for any system with finite degrees of freedom is given in the
Appendix~A. Within this model we show that the quantum effects do not
lead to an acceleration of the expansion of the universe. Finally, in
the Appendix~B, we collect the technical details of the invariant
basis for Bianchi IX spacetimes.

\section{Quantum Geometrodynamics -- the General Theory}

We start with the Hilbert-Einstein action for general relativity
in its ADM or 3+1 formulation \cite{ADM62}. As usual we choose a
foliation of the 4-dimensional manifold by a one parameter
family of the spacelike hypersurfaces, $\Sigma_t$. We assume that
the spacetime we are solving for is globally hyperbolic, and that
this kind of foliation can always be accomplished. The canonical
degrees of freedom are taken to be the six components of the
spatial metric $h_{ij}$ induced on the hypersurface $\Sigma_t$ by
the full spacetime metric $g_{\mu \nu}$ restricted to act on
vectors tangent to the spatial slice. The spacetime metric tensor
then can be written as
\begin{equation}
g_{\mu \nu}=h_{\mu \nu}+\hat{n}_{\mu}\hat{n}_{\nu},
\end{equation}
where $\hat{n}^{\mu}$ are the components of the timelike unit vector
normal to the spacelike hypersurface, $\Sigma_t$, in some coordinates
$\{x^{\alpha}\}$. In what follows, the particular choice of
coordinates we use are irrelevant.  However, in order to simplify our
calculations, we assume that, at every point of the 4-dimensional
manifold, the coordinates $\{x^{i}\}$, $i=1,2,3$, define the vectors
$\{X_{i}\}$ tangent to $\Sigma_t$.  The fourth basis vector has to be
timelike but otherwise arbitrary, so that
\begin{equation}
\hat{t}^{\mu}=Nn^{\mu}+N^{\mu},
\end{equation}
where $N$ is an arbitrary function on the manifold, and $N^{\mu}$ is
the vector tangent to $\Sigma_t$. In the $\{X_{i},\hat{n}\}$ basis,
its components ($N$,$N^i$) are the familiar lapse and shift
functions. This vector defines the flow of time, and since we
leave $N$ and $N^{i}$ unspecified, it manifests the freedom of
choice for the timelike direction. With the foliation $\Sigma_t$
and basis of tangent space given, we cast the Hilbert-Einstein
action in its 3+1 form.
\begin{equation}
\label{act}
S=\frac{1}{16\pi} \int
\sqrt{h}N[{}^{(3)}R+K_{ij}K^{ij}-K^2]d^3xdt=\frac{1}{16\pi} \int
[\pi^{ij}\dot{h}_{ij}-NC^0-N_iC^i]d^3x dt,
\end{equation}
Here, $K_{ij}$ is extrinsic curvature and is defined as the Lie
derivative of the 3-dimensional metric with respect of the unit
normal vector $n$ to $\Sigma_t$, $K$ is the trace of $K_{ij}$, and $h$
is the determinant of $h_{ij}$. The Legendre transformation is
accomplished with respect to the ``time'' derivative of the spacelike
metric and is defined as, $\dot{h}_{ij} \equiv h_i^{\phantom{i}k}
h_j^{\phantom{j}k}\mathcal{L}_th_{kl}$.  The canonical momenta are then
found to be
\begin{equation}
\pi^{ij}=\frac{\delta L_G}{\delta
\dot{h}_{ij}}=\sqrt{h}(K^{ij}-Kh^{ij}),
\end{equation}
\begin{equation}
C^0=\sqrt{h}(-{}^{(3)}R+h^{-1}\pi^{ij}\pi_{ij}-\frac{1}{2}h^{-1}\pi^2),
\hbox{ and}
 \end{equation}
\begin{equation}
C^j=-2\sqrt{h}D_i(h^{-1/2}\pi^{ij}).
 \end{equation}
By varying (\ref{act}) with respect to $N$ and $N^i$, we get the
constraints of the theory which are just $C^\alpha$.  By varying
the action with respect to the other canonical variables and their
conjugate momenta, we arrive at Hamilton equation of motions,
\begin{equation} \label{eq:hdot}
\dot{h}_{ij}=\frac{\delta H_G}{\delta
\pi^{ij}}=2h^{-1/2}N(\pi_{ij}-\frac{1}{2}h_{ij}\pi)+2D_{(i}N_{j)},
\end{equation}
\begin{equation}
\dot{\pi}^{ij}=-\frac{\delta H_G}{\delta h_{ij}}.
\end{equation}
Where $H_G=\int (NC^0+N_iC^i)d^3x$ is Hamiltonian of general relativity.

Equation (\ref{eq:hdot}) is just a definition of the momenta expressed
in terms of the ``time'' derivative of the metric of the 3-dimensional
hypersurface. It is a good place to comment on the latter. In most of
the sets of basis vectors used in practice, this definition of the
time derivative, which is based on the concept of the Lie derivative,
becomes just a partial derivative with respect to the coordinate time,
$t$. The conditions for this is that $\hat{t}$ commutes with the
spacelike vectors, ${X^i}$, of the basis. In other words it must be
coordinated with respect to the spatial basis. This condition is
trivially satisfied in the case of the basis defined by the coordinate
functions $\{t,x^i\}$ on the manifold, but one has to make sure that
it still holds in case one decides to work in some other basis, say
invariant basis as for Bianchi IX model we consider here.

When we have the classical theory cast in its canonical form, it can
be quantized. The key idea of the new formalism is to split the
classical canonical degrees of freedom, which are the six components
of $h_{ij}$, into the so called true dynamical variables ($\{q_d\}$,
$d=1-2$) and the embedding variables (${q_e}$, $e=1-4$).  Moreover, we
treat only the true dynamical variables $\{q_d\}$ as the genuine
physical degrees of freedom of the gravitational field, and as such
are susceptible to quantum fluctuations.  Whereas, the set of four
embedding variables are thought of as classical parameters of the
theory, and responsible for the appearance of the coherent spacetime
picture.  Following York's analysis of canonical structure of general
relativity, we identify the physical degrees of freedom with the
conformal 3-geometry of the spatial slice.\cite{Yor72} Next, we define
the dynamical Hamiltonian of the theory
$H_{dyn}(p^d,q_d,q_e,\dot{q}_e)$ by performing the Legendre
transformation with respect to the true dynamical variables
only. Consequently, the standard canonical methods of quantization
lead us to a Schr\"odinger equation for the functional $\Psi(q_d)$,
where in the expression for $H_{dyn}$, $q_d$ and $p^d$ becomes linear
operators with the standard canonical commutation relations.
\begin{equation}
-i\hbar\frac{\delta \Psi(q_d)}{\delta
t}=H_{Dyn}(\hat{p}^d,\hat{q}_d,q_e,\dot{q}_e)\Psi(q_d)
\end{equation}

To make the system mathematically complete, we specify the values of
the embedding variables on each $\Sigma_t$.  This is achieved by
imposing the constraints of the theory as expectation value equations.
\begin{equation}
\langle C^i(\hat{p}^d,\hat{q_d},q_e,\dot{q}_e) \rangle=0
\end{equation}
\begin{equation}
\langle C^0(\hat{p}^d,\hat{q_d},q_e,\dot{q}_e) \rangle=0
\end{equation}

This gives us a set of four first-order differential equations for
the four unknown embedding variables.  Note that these equations
do not impose restrictions on the Hilbert space of the system, but
instead can be thought of as an implicit definition of the embedding
parameters.

Therefore, in QG the time-dependent Schr\"odinger equation for the
state of the gravitational field is coupled to the set of
expectation-value constraint equations. This feature of the theory
seems to be as intriguing as it is unavoidable, and must result in
observable effects when applied to a realistic problem.

\section{Bianchi IX Cosmology}

In this section, we introduce the notation and conventions we
use to describe the homogeneous Bianchi IX cosmological model.

The characteristic feature of the Bianchi IX model is the
existence of the simply transitive isometry group \cite{RS75}. The
infinitesimal generators of this group are the three linearly
independent spacelike killing vectors, $\xi_{i}$, which obey
\begin{equation}
 [\xi_{i},\xi_{j}]=C^{k}_{\phantom{k}ij}\xi_{k},
\end{equation}
where structure constants are
$C^{k}_{\phantom{k}ij}=\varepsilon_{kij}.$

The typical orbit for this group of a point of the
four-dimensional manifold, is a spacelike hypersurface. One can 
fill the entire space with a one-parameter family of such
hypersurfaces ($\Sigma_{t}$), so that the manifold can be represented
as a direct product of the real line and this family of
hypersurfaces, $\Sigma_{t}\times R$. We use the parameter
which labels hypersurfaces as the fourth coordinate together with
other three that covers the spacelike sections. For the
latter it is convenient to use the coordinates of the three-sphere
$0\leq\theta\leq\pi$, $0\leq\varphi<2\pi$, $0\leq\psi<4\pi$ due to
the fact that the isometry group of this model is isomorphic to
the 3-dimensional rotation group.

Given the Killing vectors $\{\xi_{i}\}$ in these coordinates, one can
generate the so-called invariant basis $\{\hat{t},e_{i}\}$.  The
details of this procedure can be found in Appendix~B. In this new
basis the metric is
\begin{equation}
g=-(N^2-N^rN_r){\sigma^0}^2+N_i\sigma^0 \sigma^j+h_{ij}(t)\sigma^i
\sigma^j,
\end{equation}
where $N_i=h_{ij}N^j$ and ${\sigma^\alpha}$ is the dual basis of
the $\{\hat{t},e_i \}$.

The advantage of the invariant basis is that we are using the
symmetries of the spacetime to get rid of the explicit dependence
of the metric on spatial coordinates. In what follows, we work
in this more general invariant basis in order to keep the lapse
and shift functions ($N(t)$ and $N^i(t)$) unspecified. That
allows us to mimic the analogous degrees of freedom in the full
Hilbert-Einstein action without special symmetries. On the other
hand, we must point out that we do use the particular slicing of
the four dimensional manifold with the assigned Bianchi type IX
isometry group. This fact might not be considered a great
disadvantage from the ``3+1`` point of view. This slicing provides
a natural way to build the spacetime possessing the required
symmetries as a time evolution of a 3-dimensional space.

The metric on the spatial slices, $\Sigma_t$, of the vacuum
Bianchi IX model can be parameterized in the following manner:
\begin{equation}
h_{ij}=R_0^2e^{2\Omega}\left(e^{2\beta}\right)_{ij}\sigma^i\sigma^j;
\end{equation}
where $R_0$ is initial radius of the universe, and all other
parameters are functions of time only. The matrix
\begin{equation}
\beta_{ij}=diag[\beta_{+}+\sqrt{3}\beta_{-},\beta_{+}-\sqrt{3}\beta_{-},-2\beta_{+}],
\end{equation}
with the property $Tr\beta=0$, ensures that the 3-volume of the
hypersurface depends only on the conformal factor $\Omega$.  In
particular the 3-volume of the universe is given by $V_{universe}\sim
R_{0}^3e^{3\Omega}$.

With the use of the metric, one can calculate the 3-dimensional
quantities necessary for building the action.
\begin{equation}
 {}^{(3)}R=\frac{3}{2{R_0}^2}(1-V(\beta_{+},\beta_{-}))
\end{equation}
\begin{eqnarray}
 V(\beta_{+},\beta_{-})& = & \frac{1}{3}Tr\left( e^{4\beta}-2e^{-2\beta}+1\right) =
 \frac{2}{3}e^{4\beta_{+}}(\cosh
 4\sqrt{3}\beta_{-}-1) + 1 \nonumber\\
 & & - \frac{4}{3}e^{-2\beta_{+}} \cosh
 2\sqrt{3}\beta_{-} + \frac{1}{3}e^{-8\beta_{+}}
\end{eqnarray}
\begin{equation}
 \sqrt{h}=R_0^3\ e^{3\Omega}
\end{equation}

Taking into consideration that all the canonical variables, including
the lapse and shift functions, are independent of the spatial coordinates,
and that this model satisfies the momentum constraints identically,
we get
\begin{equation}
 S=\pi \int dt\left[ \pi^{ij} \frac{\partial h_{ij}}{\partial t} -
 NC^0 \right]
 \end{equation}
Here we have used the fact that $\int \sigma^1 \wedge \sigma^2 \wedge
\sigma^3=(4\pi)^2$. By using expressions for the momenta we
derive the first Hamilton equation.
\begin{equation}
 \frac{\partial h_{ij}}{\partial
 t}=\frac{2N}{\sqrt{h}}(\pi_{ij}-\frac{1}{2}h_{ij}\pi^k_{\phantom{k}k}),
\end{equation}
This can be solved for the momenta in terms of time derivative of
the metric.
\begin{equation}
 \pi_{ij}=\frac{\sqrt{h}}{2N}(\dot{h}_{ij}-h_{ij}\dot{h}^k_{\phantom{k}k})
\end{equation}
Here, $\dot{h}^k_{\phantom{k}k}=h^{ij}\dot{h}_{ij}$.

The special slicing and convenient choice of basis we adopt has
allowed us to reduce our problem to the dynamics of a system with only
a finite number degrees of freedom.  We further simplify the model
by substituting the particular form of metric we have chosen in to the
general expressions.
\begin{equation}
\frac{\partial h_{ij}}{\partial t} = 2\dot{\Omega}h_{ij}+h_{ij}
\dot{\beta}_{ij} \phantom{longspace}             \hbox{(no
summation!)}
\end{equation}
\begin{equation}
\dot{h}^k_{\phantom{k}k}=6\dot{\Omega} \
\end{equation}
\begin{equation}
\pi^{ij} \frac{\partial h_{ij}}{\partial
t}=12\frac{R_0^3e^{3\Omega}}{N}(-{\dot{\Omega}}^2 +
{\dot{\beta_{+}}}^2 + {\dot{\beta_{-}}}^2)
\end{equation}
\begin{equation}
C^0=R_0^3e^{3\Omega}\left( -{}^{(3)}R+
\frac{6}{N^2}(-{\dot{\Omega}}^2 + {\dot{\beta_{+}}}^2 +
{\dot{\beta_{-}}}^2)\right)
\end{equation}
Putting the two last terms together we get,
\begin{equation} \label{eq:BIXaction}
S=\pi \int
\frac{R_0^3e^{3\Omega}}{N}[-6{\dot{\Omega}}^2+6{\dot{\beta_{+}}}^2+6{\dot{\beta_{-}}}^2+N^{2}{}^{(3)}R]dt.
\end{equation}

In principle, this action (\ref{eq:BIXaction}) might be the starting
point of the theory for the Bianchi IX cosmology, but then the
relation to the full theory would be obscured. In the next three
sections we apply (1) the ADM square root, (2) the Wheeler-DeWitt
and, (3) the QG quantization schemes to the system described by this
action, respectively.

\section{ADM Square Root Quantization}

The main idea of the ADM quantization scheme is to solve the
constraints before quantizing, and to use four out of the six
canonical variables as coordinate labels. The action for the
theory, which must be quantized, become
\begin{equation}
S_{ADM}=\frac{1}{16\pi}\int d^4x\ \pi^{ij} \dot{h}_{ij}.
\end{equation}
The ADM action for the Bianchi IX cosmological model is
\begin{equation}
S_{ADM}=\int p_{+}d\dot{\beta}_{+} + p_{-}d\dot{\beta}_{-} -
\frac{m}{N}\dot{\Omega} d\Omega.
\end{equation}
In the expression above we got rid of the explicit dependence on the
coordinate time. Following the approach of Misner,\cite{Mis69} one
treats $\Omega$ as the time parameter for the evolution of the
conformal 3-geometry, then by definition, the term in the
action in front of $d \Omega$ is the ADM Hamiltonian for the
system. According to this prescription, one solves the
constraint equation to find the expression for the Hamiltonian in
terms of the canonical variables and their conjugate momenta. Recall
that in our case, the constraint equation might be rewritten as
follows:
\begin{equation}
\frac{m}{N}\dot{\Omega}=p_{+}^2 + p_{-}^2 -
\frac{m^2}{6}{}^{(3)}R;
\end{equation}
hence,
\begin{equation}
H_{ADM}^2=p_{+}^2 + p_{-}^2 - \frac{m^2}{6}{}^{(3)}R.
\end{equation}
All we need now is to take the square root of this expression
to obtain the generator for time evolution.

This last step is not problematic if one analyzes the classical
dynamics, because the magnitude of Hamiltonian
($H_{ADM}=\frac{m}{N}\dot{\Omega}$) is always a real number by
definition. Then the constraint equation puts restrictions on the initial
momenta.

The quantum mechanical system is different. Here we must define the
operator, $H_{ADM}$. This can be done if we find all of the
eigenvalues, $E_n$, and corresponding eigenvectors, $|\varphi_n
\rangle$, of $H_{ADM}^2$. If all the eignevalues are positive
($E_{n}>0$), then $H_{ADM}=\sum \sqrt{E_n} |\varphi_n \rangle \langle
\varphi_n|$. In our case, however, not all of the eigenvalues are
positive because the expression for $H_{ADM}^2$ is not positive
definite. To see that this is the case, let us set up a Gaussian state
around $\beta_{+}=0, \beta_{-}=0$ with zero initial momenta and
calculate the expectation value of the $H_{ADM}^2$ at $\Omega=0$. It
is easy to see that $\langle{}^{(3)}R\rangle=\frac{3}{2R_0^2}$ which
means that $\langle H_{ADM}^2 \rangle =
-\frac{m^2}{6}\frac{3}{2R_0^2}$ is negative. Obviously it is
impossible to get a negative expectation value of the operator which
possesses only positive eigenvalues. This shows that Hamiltonian
operator is not Hermitian on its full domain and therefore the
dynamics generated by this operator will not be unitary.

\section{Wheeler-DeWitt Equation}

In contrast to the ADM procedure,\cite{Mis69} the canonical theory of
quantum gravity suggested by DeWitt \cite{DeW67} does not try to
reduce the set of canonical variables by solving constraint
equations. Instead it treats all of the components of the spatial
metric on equal footing by promoting them and their conjugate momenta
to the linear operators acting on the Hilbert space of physical
states. The latter is defined by the requirement that the physically
admissible states have to be the eigenvectors of the constraint
operators with zero eigenvalues. So the constraints are realized as
restrictions on the Hilbert space of the states of the theory. This
approach was first introduced by P.A.M.~Dirac as a general method of
canonical quantization of theories with first-class
constraints.\cite{Dir64}  Einstein's theory of general relativity
falls in this category, so it is logical to apply  the general
prescription of Dirac to it. The first non-trivial step in the
procedure is to determine the physical states of the theory.  In
the case of gravity, they are the solutions of
\begin{equation}
\hat{C}^i\Psi_{phys}=0, \hbox{ and}
\end{equation}
\begin{equation}
\label{diffcon}
\hat{C}^0\Psi_{phys}=0.
\end{equation}
The first condition ensures that the wave-functional,
$\Psi(h_{ij})$, depends only on the 3-geometry of the spatial
hypersurface. It can be realized by choosing the appropriate
coordinate-independent representation for the wave-functional. For
example, one can require $\Psi$ to be a function of the powers of
scalar 3-curvature (${}^{(3)}R$). Whichever way one chooses to
implement this condition, it is not expected to give us any
dynamics, simply because it depends on the state of the field on
the spatial slice and gives one no information on what the
physical state must be off of the slice. It does not map the physical
states on different $\Sigma_t$ hypersurfaces.

The idea of DeWitt was that the last constraint equation
(\ref{diffcon}), which is quadratic in canonical momenta, ought to
serve this purpose.  One assumes that this equation, by itself,
contains all the dynamics of the gravitational field. The classical
expression for this constraint can be rewritten in the following way:
\begin{equation}
G_{ijkl}\pi^{ij}\pi^{kl}-\sqrt{h}{}^{(3)}R=0;
\end{equation}
where $G_{ijkl}=\frac{1}{2}h^{-1/2}\left(
h_{ik}h_{jl}+h_{il}h_{jk}-h_{ij}h_{kl} \right)$ is the DeWitt's
supermetric. It is the metric on the space of 3-metrics which
is a six dimensional manifold with $h_{ij}$ being six independent
components of a tangent vector. Note that this notation differs
from the standard one by the use of the lower indices instead of
upper ones. In all other respects one can fully rely on the
differential geometry methods applied to the six dimensional
manifold and find the inverse supermetric,
\begin{equation}
G^{ijkl}=\frac{1}{2}h^{1/2}\left(
h^{ik}h^{jl}+h^{il}h^{jk}-2h^{ij}h^{kl} \right),
\end{equation} 
which can be used to contract two vectors. An interesting feature of
this construction is that $G_{ijkl}$ has signature $[-+++++]$, which
means that it induces a time-like direction on the space of
3-metrics. This fact supports DeWitt's expectation that the
Hamiltonian constraint possesses the dynamical content of the theory.

In the quantum theory, with the particular coordinate representation
we have chosen, the canonical momenta operators become functional
differential operators,
\begin{equation}
\pi^{ij}=-i\frac{\delta}{\delta h_{ij}}.
\end{equation}
The constraint equation is a Klein-Gordon-type equation,
\begin{equation} \label{eq:WDW}
\left( G_{ijkl} \frac{\delta}{\delta h_{ij}} \frac{\delta}{\delta
h_{kl}} + \sqrt{h}{}^{(3)}R \right)=0,
\end{equation}
where the first term due to the signature of the metric, and is a kind
of Laplace-Beltrami operator in a six-dimensional Riemannian
manifold, with $G_{ijkl}$ being its contravariant metric. Thus we
have Wheeler-DeWitt equation (\ref{eq:WDW}).

From the theory of Klein-Gordon equation, it is known that the
space of solutions can be assigned with a naturally conserved
inner product. The inner product for Wheeler-DeWitt equation takes
the form,
\begin{equation} \label{eq:WDWproduct}
(\Psi_a,\Psi_b)=Z\int_{\Sigma}\prod_x -i\left(
\Psi_{a}^*G_{ijkl}\frac{\delta\Psi_{b}}{\delta
h_{kl}}-\frac{\delta \Psi_{a}^*}{\delta h_{kl}}G_{ijkl}\Psi_{b}
\right) d\Sigma^{ij}.
\end{equation}
Here $Z$ is a proper normalization factor, the integral is taken
in the space of metrics over a product of a set of
five-dimensional hypersurfaces $\Sigma(x)$ defined for each point
of three-dimensional spatial slice, and $d\Sigma_{ij}$ is the
directed volume element for each five-dimensional hypersurface
respectively. This inner product is explicitly independent of the
choice one makes for the $\Sigma(x)$ at each point of the spatial
slice.

These general expressions, (\ref{eq:WDW}) and
(\ref{eq:WDWproduct}), of the Wheeler-DeWitt theory, when applied
to the particular case of homogeneous cosmologies, are simplified
tremendously. The main reduction comes from the independence of
the 3-metric on the spatial coordinates. Thus, instead of the
infinite product over all points of 3-dimensional spatial
slice, we have just one integration in the space of metrics for
all points. In addition, the functional differential operators
become simple partial derivatives.

In case of the Bianchi IX metric, the 3-metric $h_{ij}$ is diagonal
and therefore is determined by only three independent parameters. In
the particular parameterization of the metric we use, the
Hamiltonian constraint is,
\begin{equation} \label{eq:WDWconstraint}
C^0=\frac{1}{2m} \left( -p_{\Omega}^2+p_{+}^2+p_{-}^2 -
\frac{m^2}{6} {}^{(3)}R \right)=0,
\end{equation}
where $p_{\Omega}=-\frac{m}{N}\dot{\Omega}$ is the canonical momenta
conjugate to $\Omega$.  Note that we treat all of the variables on the
same footing, and therefore, we have to introduce the momenta for
conformal factor.  In ridding ourselves of overall factor
$\frac{1}{2m}$, we obtain the Wheeler-DeWitt equation.
\begin{equation}
\left(-\frac{\partial^2}{\partial \Omega^2} +
\frac{\partial^2}{\partial \beta_{+}^2} +
\frac{\partial^2}{\partial \beta_{-}^2} + \frac{m^2}{6} {}^{(3)}R
\right) \Psi(\Omega,\beta_{+},\beta_{-})=0
\end{equation}

The coordinate basis $\{ \frac{\partial}{\partial \Omega},
\frac{\partial}{\partial \beta_{+}}, \frac{\partial}{\partial
\beta_{-}}\}$ is orthogonal, although it is not an orthonormal
basis with respect to the supermetric. If we choose the
hypersurface in the space of the metric as the surface of constant
$\Omega$, then up to irrelevant normalization factor, the inner
product for the solutions of Wheeler-DeWitt equation become,
\begin{equation} \label{eq:BIXproduct}
(\Psi_a,\Psi_b)=-i \int \left( \Psi_a^* \frac{\partial
\Psi_b}{\partial \Omega} - \frac{\partial \Psi_a^*}{\partial
\Omega} \Psi_b \right) d\beta_{+} d\beta_{-}.
\end{equation}
At this point one just needs to solve the resulting equation, and
by specifying the appropriate boundary conditions, choose among all
possible solutions those which represents the physical situation
in question.

It is this last step of the procedure that gives rise to a major flaw
of the theory. In particular, the natural product (\ref{eq:BIXproduct}) is not
positive definite for a generic solution of the Wheeler-DeWitt
equation.  Therefore, if one is to give probabilistic interpretation to
wave-function, $\Psi(\Omega,\beta_{+},\beta_{-})$, one will have to
restrict the space of the solutions to those that have a positive
norm, they would be positive frequency solutions with respect some
natural notion of time. For a general spacetime, it is ordinarily
impossible to accomplish such a subdivision of solutions into positive
and negative frequency solutions due to the lack of symmetries. One is
faced with the dilemma to either (1) assign a different inner product
to the wave-function or even turn it into a field and accomplish third
quantization, or (2) to restrict ones attention to a particular class
of physically relevant spacetimes whose symmetries or asymptotic
behavior allow such a division and impose a boundary condition which
generates it.  Despite many efforts, neither of the two paths has led to
a consistent and general quantum theory of gravity.

\section{Dynamical Hamiltonian and the Schr\"odinger Equation}

Having the action for the system at our disposal (\ref{eq:BIXaction}),
we proceed with the standard canonical procedure to obtain
Hamilton's equations of motion. We deviate here from the textbook
prescription and perform a Legendre transformation of the Lagrangian only
with respect to the variables $\beta_{+}$ and $\beta_{-}$. This
promotes the two anisotropy variables as the parameters that represent
the true dynamical degrees of freedom of the spacetime.  We
determine the momenta conjugate to $\beta_{+}$ and $\beta_{-}$.
\begin{equation} \label{pplus}
p_{+}=\frac{\partial L_G}{\partial \dot{\beta_{+}}}=\frac{12\pi
R_0^3e^{3\Omega}}{N}\dot{\beta_{+}}=\frac{m}{N}\dot{\beta_{+}}
\end{equation}

\begin{equation} \label{pminus}
p_{-}=\frac{\partial L_G}{\partial \dot{\beta_{-}}}=\frac{12\pi
R_0^3e^{3\Omega}}{N}\dot{\beta_{-}}=\frac{m}{N}\dot{\beta_{-}}
\end{equation}

We define the dynamical Hamiltonian as
\begin{equation}
H_{Dyn}=p_{+}\dot{\beta_{+}}+p_{-}\dot{\beta_{-}}-L_G=N\left(\frac{p_{+}^2}{2m}+
\frac{p_{-}^2}{2m} \right) + \frac{m}{2N}\left( {\dot{\Omega}}^2-
\frac{N^2}{6}{}^{(3)}R \right).
\end{equation}
The action, written in terms of the new variables, is thus
\begin{equation} \label{eq:BIXactionhamilton}
S=\int
\left(p_{+}\dot{\beta_{+}}+p_{-}\dot{\beta_{-}}-H_{Dyn}(p_+,p_-,\beta_+,\beta_-,\dot{\Omega},\Omega,N)
\right) dt.
\end{equation}
According to the principle of extremal action we get Hamilton's
equation of motion for the true dynamical variables by varying
(\ref{eq:BIXactionhamilton}) with respect to $\beta_{+}$, $\beta_{-}$
and their conjugate momenta (\ref{pplus},\ref{pminus}). Also, we get
one constraint equation from the variation with respect to the lapse,
$N$.  The equation of motion for the conformal factor is redundant. We
do not consider it as a dynamical degree of freedom, but merely as a
time dependent parameter.

The constraint equation becomes
\begin{equation}
p_{+}^2+p_{-}^2=\frac{m^2}{N^2}\left(
{\dot{\Omega}}^2+\frac{N^2}{6}{}^{(3)}R \right).
\end{equation}

We are now in the position to quantize our theory. The canonical
variables are the two anisotropy parameters, $\beta_{+}$ and
$\beta_{-}$, with their conjugate momenta and standard commutation
relations. The conformal factor, $\Omega$, is treated as a c-number
function that depends only on time. The constraint is imposed as an
expectation-value equation, which in effect determines the time
dependence of the conformal factor. The evolution of the quantum state
is generated by the dynamical Hamiltonian, $H_{Dyn}$, which results in
a Schr\"odinger equation for the wave function
$\psi(\beta_{+},\beta_{-},t)$.
\begin{equation} \label{eq:Schrodinger}
-i\hbar\frac{\partial \psi}{\partial
t}=H_{Dyn}\psi=\left[\frac{N}{2m}\left( \frac{\partial^2}{\partial
\beta_{+}} + \frac{\partial^2}{\partial \beta_{-}} \right) +
\frac{m}{2N}\left( {\dot{\Omega}}^2-\frac{N^2}{6}{}^{(3)}R\right)
\right] \psi
\end{equation}
The constraint equation produces a first order equation
for $\Omega$
\begin{equation} \label{eq:constraint}
\frac{m^2}{N^2}{\dot{\Omega}}^2 = \langle p_{+}^2+p_{-}^2 -
\frac{m^2}{6}{}^{(3)}R \rangle
\end{equation}

We obtain a system of coupled integro-differential equations,
(\ref{eq:Schrodinger}) and (\ref{eq:constraint}), which exhibit an  explicit
dependence on the lapse, $N$. One requirement of any quantum theory of
gravity is that measurable predictions should be independent of the
choice of lapse and shift functions, since different choices, after
all, just represents different choices of basis. Shift functions have
never appeared in this model, because the momentum constraints were
satisfied identically, but lapse function is present. It is important
to confirm that the system of equations given above give rise to dynamics
which is independent of the choice $N$, despite its explicit
appearance. To show this lapse invariance, let us note that, since
the conformal factor is just a real function of coordinate time
$\Omega(t)$, one can find the inverse function (we assume that it is
always can be done) $t(\Omega)$, and use $\Omega$ as the time parameter
of the theory. In that case our wave function becomes
$\psi(\beta_{+},\beta_{-},\Omega)$.  Furthermore, using
\begin{equation}
\frac{\partial \psi}{\partial t} = \frac{\partial \psi}{\partial
\Omega}\frac{\partial \Omega}{\partial
t}=\dot{\Omega}\frac{\partial \psi}{\partial \Omega}.
\end{equation}
we solve the constraint equation (\ref{eq:constraint}) for
$\dot{\Omega}$.
\begin{equation} \label{eq:omegadot}
{\dot{\Omega}}^2=\frac{N^2}{m^2}\langle \tilde{C}_0 \rangle
\end{equation}
Here, $\tilde{C}_0=p_{+}^2+p_{-}^2 - \frac{m^2}{6}{}^{(3)}R$. In
the last step we eliminate $\dot{\Omega}$ from the equation
for the wave function with the use of (\ref{eq:omegadot}). After
simplification we obtain,
\begin{equation}
-i \hbar \frac{\partial \psi(\beta_{+},\beta_{-},\Omega)}{\partial
\Omega} = \frac{1}{2\sqrt{\langle \tilde{C}_0 \rangle}}
\left[\frac{\partial^2}{\partial \beta_{+}} +
\frac{\partial^2}{\partial \beta_{-}} + m^2 \left( \frac{\langle
\tilde{C}_0 \rangle}{m^2} - \frac{{}^{(3)}R}{6} \right) \right]
\psi(\beta_{+},\beta_{-},\Omega).
\end{equation}

This integro-differential equation is explicitly independent of
the choice of the lapse function. In practice it is more
convenient to deal with the system of the coupled equations where
one can make any choice for $N(t)$, for example, we take $N(t)=1$.

Another requirement of a reasonable candidate for a quantum theory
is the existence of the classical limit. Basically that means that
the theory should posses a regime which would mimic the
results of the classical theory. The dynamics of the theory on the
quantum level is governed by $\hat{H}_{Dyn}$. It is natural for one
ask if the dynamics of the classical theory can be
generated by the same Hamiltonian.  Recall that the $H_{Dyn}$ is
defined prior the quantization by the standard Legendre
transformation of the Lagrangian of the full theory with respect
of the reduced set of dynamical variables. It is therefore logical
to expect that if this Hamiltonian accomplishes the time
translation on the quantum level it will do so for the classical
theory. It is not intuitively obvious that it is true for general
relativity. It is certainly not the case for a general mechanical
system, so the question is if there is such class of theories for
which it is possible to reproduce the canonical equations of
motion by defining the Hamiltonian over the reduced set of
variables.

In the case under consideration, the full canonical Hamiltonian,
sometimes called super-Hamiltonian, is $H_{G}=NC^0$ with the
constraint given by (\ref{eq:WDWconstraint}). The canonical
equations of motion are
\begin{equation}
\dot{q}_a=\{q_a,H_G\}, \hbox{ and}
\end{equation}
\begin{equation}
\dot{p}^a=\{p^a,H_G\},
\end{equation}
which subject to the constraint,
\begin{equation}
C^0(q_a,p^a)=0.
\end{equation}
Here, $\{q_a\}$ stands for $\{\Omega,\beta_+,\beta_-\}$.

On the other hand, the equations of motion under the dynamical
Hamiltonian, $H_{Dyn}$, are
\begin{equation}
\dot{q}_d=\{q_d,H_{Dyn}\}, \hbox{ and}
\end{equation}
\begin{equation}
\dot{p}^d=\{p^a,H_{Dyn}\},
\end{equation}
and are subject to the constraint,
\begin{equation}
C^0(q_d,p^d,q_e,\dot{q}_e)=0.
\end{equation}
In this case,  $\{q_d\}=(\beta_+,\beta_-)$ and $q_e \equiv \Omega$
It is shown (Appendix~A) that the two sets of equations are
equivalent.

Actually we have proved the more general fact, that for a system of
$N$ degrees of freedom described by a Hamiltonian of the type
$H=N_{\alpha}C^{\alpha}$, where $C^{\alpha}=0$ are the $N_c < N$
constraints of the first class in Dirac's terminology, one can
define the reduced Hamiltonian over $N_d=N-N_c$ degrees of
freedom. Moreover the canonical equations of motion generated by
this reduced Hamiltonian, together with the constraint equations,
are equivalent to the full canonical theory. In turn, this
means that, under these conditions, the constraints of  the theory
give rise to the equations of motion for the $N_c$ degrees of
freedom. A detailed proof of this can be found in 
Appendix~A. Here we just emphasize that the proof relies heavily
on the fact that the constraints are of the first class. It seems
almost trivial to generalize this proof to the case of field
theories of the same type, therefore the statement is true
even in the case of full theory of general relativity.

Now that we have established that the $H_{Dyn}$ works well as the
Hamiltonian of the classical theory, we just need to invoke
Erenfest's theorem.  It is best formulated in the Heisenberg
picture where time dependence of the operators are given by,
\begin{equation}
\dot{\hat{q}}_d=-\frac{i}{\hbar}[\hat{q}_d,\hat{H}_{Dyn}], \hbox{
and}
\end{equation}
\begin{equation}
\dot{\hat{p}}^d=-\frac{i}{\hbar}[\hat{p}^d,\hat{H}_{Dyn}].
\end{equation}
Based on the mapping between Poisson brackets and commutator
established by $\{{},{}\} \to i\hbar[{},{}]$ we see that, at least
on the level of algebra, the Heisenberg equations of motion
reproduce the classical theory. Therefore if we consider the
states that are peaked around the classically allowed values  of
the canonical variables with small dispersion then
\begin{equation}
\langle \dot{\hat{q}}_d \rangle=-\frac{i}{\hbar}\langle
[\hat{q}_d,\hat{H}_{Dyn}] \rangle,
\end{equation}
\begin{equation}
\langle \dot{\hat{p}}^d \rangle=-\frac{i}{\hbar}\langle
[\hat{p}^d,\hat{H}_{Dyn}] \rangle,
\end{equation}
together with
\begin{equation}
\langle C^0(\hat{q}_d,\hat{p}^d,q_e,\dot{q}_e) \rangle=0,
\end{equation}
is guaranteed to give us the desired classical limit of the
theory.

As we have shown above, the states that can be called semi-classical
are those that are highly peaked around a classically allowed
configuration of canonical variables.  Furthermore, such states will
follow the classical trajectories in the phase space. However, are
there new behaviors exhibited by those states that are far from being
classical? The characteristic feature of the resulting quantum theory
of the Bianchi~IX cosmology is the constraint equation that is
coupled to the Schr\"odinger equation. This equation governs the
evolution of conformal factor, $\Omega(t)$. Although $\Omega(t)$ is a
classical variable with no dispersion allowed; nevertheless, it can
deviate from the classical trajectory due to the coupling to the
quantum state, $\psi(\beta_+,\beta_-,t)$, of the system. For highly
non-classical wave functions, one can expect to find substantial
deviations or even qualitatively different behavior. It is known that
the Bianchi~IX cosmology does not predict a universe expanding with an
acceleration. Now in the light of this quantum theory we readdress
this issue.

It is easy to show that the quantum theory developed here is
consistent with the deceleration behavior of the classical model
in that it predict no accelerated expansion of the universe for
any state of the system. This result is independent of the degree
of non-classicity of the initial state function. To demonstrate
this, we calculate explicitly the time derivative of the
constraint.
\begin{equation}
\frac{d \langle \hat{C}^0 \rangle}{dt}= -\frac{i}{\hbar} \langle
[\hat{C}^0,\hat{H}_{Dyn}] \rangle + \frac{\partial \langle
\hat{C}^0 \rangle}{\partial t}=0
\end{equation}
The commutator of the constraint with $H_{Dyn}$ is zero. What we
are left with is just the partial derivative with respect to time.
It is convenient for our purposes to use the constraint equations
in the form,
\begin{equation}
\dot{\Omega}^2=\frac{N^2}{m^2} \langle \hat{p}_+^2+\hat{p}_-^2
\rangle -\frac{N^2}{6} \langle {}^{(3)}\hat{R} \rangle.
\end{equation}
Recalling the definition of the effective ``mass'' $\left(m=12\pi
R_0^3e^{3\Omega}\right)$, we find the time derivative of the
constraint.
\begin{equation}
\ddot{\Omega}=-3\frac{N^2}{m^2}\langle \hat{p}_+^2+\hat{p}_-^2
\rangle
\end{equation}
This explicitly shows that $\ddot{\Omega}<0$; therefore, the universe
always decelerates.

\section{Conclusions}

The theory of canonical quantum gravity (Quantum Geometrodynamics)
presented here has been applied to the homogeneous Bianchi IX
cosmology. It has been shown that the general formalism leads to a
well defined and consistent quantum theory, which unlike other
approaches outlined here (ADM and Wheeler-DeWitt), allows one to make
predictions relevant for cosmology. It has been proven that the
equation of QG is independent of the choice of the time variable,
(they are independent of the lapse function), and that in the
classical limit, they agree with the classical Einstein's equations of
motion. This latter fact seems to be generalizable to the full field
theoretic case. The constraints are imposed as expectation-value
equations and provide the coupling of quantum dynamical variables with
the classical embedding parameters which, in turn,  are responsible for the
appearance of the coherent classical spacetime. This way, the
constraints of the classical theory have been given a new
interpretation and physical meaning. On the practical side, the theory
consists of a Schr\"odinger equation coupled to the first-order
differential expectation valued equation. This system of equations can
be successfully solved numerically, which was done in the case of
various cosmologies including the Bianchi IX cosmology
\cite{KheMil96,KheMil00}. It was originally expected that one
might find effects different from the predictions of the classical
theory for the highly non-classical states, for example accelerated
expansion of the universe. At least for Bianchi IX model we have
analytically established that his cosmological acceleration does not
occur. The expansion of this model universe always decelerates.

\section{Acknowledgements}

We thank Chris Beetle for many stimulating
discussions on this and related topics.  In addition to helping
forward these calculations, these discussions have further reinforced
our earlier perceptions that we have much more to explore within this
QG approach. We wish to thank the Charles E. Schmidt College of
Science and the Florida Atlantic University Office of Research for
their support.

\section{Appendix A: A Proof of Equivalency of the Two Dynamical Pictures.}

For the sake of generality, consider a system with $N$ degrees of
freedom subjected to $N_c < N$ constraints of the first class with the
full canonical Hamiltonian of the form
\begin{equation}
H=N_iC^i(q_a,p^a),
\end{equation}
where $C^i(q_a,p^a)=0$ are the constraints of the first class.
Consequently, their Poisson brackets with each other weakly vanish,
$\{C^i,C^j\}=f^{ij}_{\phantom{ij}k} C^k$. $\{q_a\}$, $a=1 - N$ stands
for the set of canonical variables, and $N_i$, $i=1 N_c$ are the
Lagrange multipliers of the theory; the degrees of freedom that give
us the primary constraints of the theory.

Let us split the set of canonical variables, $\{q_a\}$, $a=1 - N$,
into two sets $\{q_e\}$, $e=1 - N_c$ and $\{q_d\}$, $ d=(N_c+1) \div
N$. We then define reduced Hamiltonian by performing a Legendre
transformation only over the second set of variables:
\begin{equation}
H_R=p^d \dot{q}_d(p^d,q_d)-L(q_d,\dot{q}_d(p^d,q_d),q_e,\dot{q}_e)
\end{equation}
Here the full Hamiltonian is,
\begin{equation}
H=p^d
\dot{q}_d(p^d,q_d,p^e,q_e)+p^e\dot{q}_e(p^d,q_d,p^e,q_e)-L(q_d,\dot{q}_d(p^d,q_d,p^e,q_e),q_e,\dot{q}_e(p^d,q_d,p^e,q_e)).
\end{equation}

The pair of canonical equations of motion for the dynamical
variables, $\{q_d,p^d\}$, will be the same no matter what
Hamiltonian is being used, simply because it is derived from the
variational principle of the action, and this action is not
altered by the Legendre transformation. Nevertheless, we show
here in details that they are the same.

The canonical equations of motion for $\{q_d,p^d\}$ with the full
Hamiltonian are,
\begin{equation} \label{eq:momentaH}
\dot{q}_d=\frac{\partial H}{\partial p^d}, \hbox{ and}
\end{equation}
\begin{equation}
\dot{p}_d=-\frac{\partial H}{\partial q^d}.
\end{equation}
Expressed in terms of the reduced Hamiltonian, they are,
\begin{equation} \label{eq:momentaHdyn}
\dot{q}_d=\frac{\partial H_R}{\partial p^d}, \hbox{ and}
\end{equation}
\begin{equation}
\dot{p}_d=-\frac{\partial H_R}{\partial q^d}.
\end{equation}

What we need to show is that the right hand sides of these equations
are the same. Equations (\ref{eq:momentaH}) and (\ref{eq:momentaHdyn})
just define the momenta.  They are obviously the same because the
momenta are defined with respect to the Lagrangian and not the
Hamiltonian of the system.  To see that the other equations are
equivalent, we have to calculate the right hand side.
\begin{equation}
-\frac{\partial H_R}{\partial q^d}=-p^d\frac{\partial
\dot{q}_d}{\partial q_d}+\frac{\partial L}{\partial
q_d}+\frac{\partial L}{\partial \dot{q}_d} \frac{\partial
\dot{q}_d}{\partial q_d}=\frac{\partial L}{\partial
q_d}=-\frac{\partial H}{\partial q^d}
\end{equation}
Here we have used the definition of the momenta,
$p^d=\frac{\partial L}{\partial \dot{q}_d}$.

As far as the dynamical variables ($\{q_d\}$) are concerned, it
does not matter which Hamiltonian is used. Incidentally, this leads
to the vanishing of the Poisson bracket of the full Hamiltonian,
with the reduced Hamiltonian defined on the reduced phase space.
It is important to note that the full Hamiltonian ($H$) is a
constraint of the system, and therefore its Poisson brackets
should vanish for consistency.
\begin{equation}
\{H,H_R\}_{Dyn}=\frac{\partial H}{\partial q_d}\frac{\partial
H_R}{\partial p^d}-\frac{\partial H}{\partial p^d}\frac{\partial
H_R}{\partial q_d}=0
\end{equation}

Let us shift our attention to the set of embedding variables
$\{q_e\}$. The canonical equations of motion for these are
governed by the full Hamiltonian.
\begin{equation}
\dot{q}_e=\frac{\partial H}{\partial p^e}=N_i\frac{\partial
C^i}{\partial p^e}
\end{equation}
\begin{equation}
\dot{p}_e=-\frac{\partial H}{\partial q^e}=-N_i\frac{\partial
C^i}{\partial q^e}
\end{equation}

In the picture with the reduced Hamiltonian, we expect the
constraint equations to provide us with the equivalent evolution.
The constraint equations are first order differential equations.
One has to take the time derivative of the constraints to get the
second order equations of motion.
\begin{equation}
\frac{d}{dt}C^i(q_d,p^d,q_e,p^e)=\{C^i,H_R\}_{Dyn}+\frac{\partial
C^i}{\partial t}=0
\end{equation}

We consider $q_e$ and $p^e$, not as canonical variables, but as
time-dependent functions, $p^e$ is related to $\dot{q}_e$ through this
definition. The time derivative of the constraint becomes,
\begin{equation}
\{C^i,H\}_{Dyn}+\frac{\partial C^i}{\partial p^e}\dot{p}^e +
\frac{\partial C^i}{\partial q^e}\dot{q}^e=0.
\end{equation}

Therefore we have $N_c$ equations. We solve this system of
equations for the $N_c$ time derivatives of the momenta.
\begin{equation}
\dot{p}^k=-{\left(\frac{\partial C^k}{\partial
p_i}\right)}^{-1}\left(\frac{\partial C^i}{\partial
q_e}\right)N_l\frac{\partial C^l}{\partial
p^e}-{\left(\frac{\partial C^k}{\partial
p_i}\right)}^{-1}\{C^i,H\}_{Dyn}
\end{equation}
Here, we have used the relation between the velocities and momenta
$\dot{q}_e=N_l\frac{\partial C^l}{\partial p^e}$. Note that the
first term appears to be an incomplete Poisson bracket of the $C^i$
and $C^l$ over the embedding variables. We add and subtract
the appropriate term to complete these brackets, and we write
down explicitly the content of the second term using the
Hamiltonian $H=N_lC^l$.
\begin{eqnarray}
\dot{p}^k & = &-{\left(\frac{\partial C^k}{\partial
p_i}\right)}^{-1}\left(\frac{\partial C^i}{\partial
q_e}\right)N_l\frac{\partial C^l}{\partial p^e} -
{\left(\frac{\partial C^k}{\partial p_i}\right)}^{-1}N_l
\left(\frac{\partial C^i}{\partial q_d}\frac{\partial
C^l}{\partial p^d} - \frac{\partial C^i}{\partial
p^d}\frac{\partial C^l}{\partial q_d} \right) + {}\nonumber\\
& &{}+{\left(\frac{\partial C^k}{\partial
p_i}\right)}^{-1}\left(\frac{\partial C^i}{\partial
p^e}\right)N_l\frac{\partial C^l}{\partial q_e} -
{\left(\frac{\partial C^k}{\partial
p_i}\right)}^{-1}\left(\frac{\partial C^i}{\partial
p^e}\right)N_l\frac{\partial C^l}{\partial q_e}
\end{eqnarray}
Next, we combine the first three terms to produce the Poisson
bracket of the constraints $C^i$ and $C^l$ taken over full set of
variables $\{q_e,q_d\}$.
\begin{equation}
\dot{p}^k=-{\left(\frac{\partial C^k}{\partial
p_i}\right)}^{-1}N_l\{C^i,C^l\} - {\left(\frac{\partial
C^k}{\partial p_i}\right)}^{-1}\left(\frac{\partial C^i}{\partial
p^e}\right)N_l\frac{\partial C^l}{\partial q_e}
\end{equation}
The Poisson brackets vanish in the weak sense, whereas the
second term produces a Kroneker delta.
\begin{equation}
\dot{p}^k \approx - \delta^k_eN_l\frac{\partial C^l}{\partial
q_e}=-N_l\frac{\partial C^l}{\partial q_k}=-\frac{\partial
H}{\partial q^k}
\end{equation}

This result shows that, as long as constraint equations are
satisfied for all times, they ensure the correct dynamical
behavior of the embedding variables. The constraint equations can
be used to generate the equations of motion for these variables
which are equivalent to the canonical ones.

To conclude this section, we would like to emphasize that the proof
is based on the fact that the constraints of the theory are of
the first class and the full canonical Hamiltonian of the theory
is the constraint itself. These features are typical for general
covariant systems.

\section{Appendix B: Invariant Basis for Bianchi IX Cosmology}

With the choice of coordinates we adopted, that of the three-sphere
$0\leq\theta\leq\pi$, $0\leq\varphi<2\pi$, $0\leq\psi<4\pi$, the
Killing vectors are,
\begin{equation}
\xi_1=\partial_{\varphi},
\end{equation}
\begin{equation}
\xi_2=\cos \varphi \partial_{\theta}-\cot \theta \sin \varphi
\partial_{\varphi}+\frac{\sin \varphi}{\sin \theta}
\partial_{\psi}, \hbox{ and}
\end{equation}
\begin{equation}
\xi_3=-\sin \varphi \partial_{\theta}-\cot \theta \cos \varphi
\partial_{\varphi}+\frac{\cos \varphi}{\sin \theta}
\partial_{\psi}.
\end{equation}

It is well known that the components of the metric tensor of a
spacetime which posses an isometry group take the simplest form in
the so-called invariant basis, $\{e_{\mu}\}$. The vectors of this
basis are transported by Lie differentiation with respect to
the Killing vectors. Mathematically, that means that
\begin{equation}
[\xi_{i},e_{\mu}]=0
\end{equation}

To demonstrate the properties of the components of the metric
tensor acquired in this basis, consider their Lie derivative with
respect to the Killing vector $\xi_{i}$,
\begin{eqnarray}
\mathcal{L}_{\xi_{i}}g(e_{\mu},e_{\nu})& = &\xi_{i} [g_{\mu
\nu}]=(\mathcal{L}_{\xi_{i}}g)(e_{\mu},e_{\nu})+g(\mathcal{L}_{\xi_{i}}
e_{\mu},e_{\nu})+g(e_{\mu},\mathcal{L}_{\xi_{i}}e_{\nu}) \nonumber\\
& = & g([\xi_{i},e_{\mu}],e_{\nu})+g(e_{\mu},[\xi_{i},e_{\nu}])=0
\end{eqnarray}
In this basis, the components of the metric are constants along the
curves generated by the Killing vectors of the given isometry group.
In our case, there are three linearly independent Killing vectors,
which means that the components of the metric do not depend on
the three spatial coordinates but only on the fourth one -- time.

To generate the invariant basis for the Bianchi IX spacetime we
proceed the following way. First, we choose a vector that corresponds
to the time coordinate as the timelike vector of the invariant basis,
$e_0=\partial_t$. This is possible because the Killing vectors,
$\xi_{i}$, are the same for every hypersurface,
$\Sigma_{t}$. Therefore, they are explicitly $t$-independent, and
$[\partial_t,\xi_{i}]=[\partial_t,\alpha_i^{\mu}\partial_{\mu}]=\alpha_{i,t}^{\mu}\partial_{\mu}=0$.
The same time independence can be imposed on the spacelike vectors of
the invariant basis by demanding that $[e_0,e_j]=0$. To generate the rest
of the basis we use,
\begin{equation}
[\xi_{i},e_{j}]=[\xi_{i},\epsilon_j^m \partial_m]=0,
\end{equation}
which results in a system of nine first order partial differential
equations for three unknown functions
$\epsilon_{j}^m(\theta,\varphi,\psi)$ for each vector. One can solve
these up to the choice of three free constants which corresponds to
the freedom of choice of the basis vector at the ``origin'' (any point
on the three-dimensional hypersurface). Using this freedom one can get
the invariant basis that obeys the same commutation relations as the
Killing vectors, namely
$[e_i,e_j]=\varepsilon^{k}_{\phantom{k}ij}e_k.$ We list them here
together with the dual basis of one-forms.
\begin{equation}
e_0=\partial_t
\end{equation}
\begin{equation}
e_1=-\sin \psi \partial_{\theta}+\frac{\cos \psi}{\sin \theta}
\partial_{\varphi}-\cot \theta \cos \psi \partial_{\psi}
\end{equation}
\begin{equation}
e_2=\cos \psi \partial_{\theta}+\frac{\sin \psi}{\sin \theta}
\partial_{\varphi}-\sin \psi \cot \theta \partial_{\psi}
\end{equation}
\begin{equation}
e_3=\partial_{\psi}
\end{equation}
\begin{equation}
\sigma^0=dt
\end{equation}
\begin{equation}
\sigma^1=-\sin \psi d\theta+\sin \theta \cos \psi d\varphi
\end{equation}
\begin{equation}
\sigma^2=\cos \psi d\theta+\sin \theta \sin \psi d\varphi
\end{equation}
\begin{equation}
\sigma^3=\cos \theta d\varphi+d\psi
\end{equation}
If the timelike vector was chosen to be the unit normal to the
hypersurface ($\Sigma_t$), the metric would take the following
form in this basis:
\begin{equation}
g=-dt^2+h_{ij}(t)\sigma^i \sigma^j.
\end{equation}

\end{document}